# Explosive Material Detection and Security Alert System

Salman Haider, Usman Saeed, Jawad Ashraf, Dr. Fareeha Zafar

Department of Computer Science, GC University Lahore

ravian.salman@hotmail.com, ravian.usman@hotmail.com, jawad1492@gmail.com, dr.f.zafar@gcu.edu.pk

## 1. ABSTRACT

The terrorism rate in Pakistan becomes higher even after the advancement of information technology. Especially APS Attack and numerous other in different part of country. The root cause of such attacks as according to our research is as: terrorists utilizing the benefit of lack of a full proof security check system. Traditional explosive detection systems are large in size, expensive and require manual attention. These systems are not much useful due to its public visibility intruder or terrorists can easily bypass the system using another route. This term paper is mainly focusing on explosive material detection using IoT with WSN. Explosive Material Detection and security alert system (e-DASS) consists of several hundreds of nodes depending upon the geographical area we are going to cover. Each node should be able to communicate with the other node and update the information if necessary. Tracking of the target can be done in an easier and faster way because all the nodes are synchronized. (e-DASS) is a power efficient explosive detection system. Most of the times nodes will be in the idle state, unless and until positive presence of an explosive is found.

## 2. INTRODUCTION

A wireless sensor network contains thousands of nodes that are dispensed in a random way, and The nodes have tendency to communicate with each other and can take decisions that are based on the sensor data. A wireless sensor network has multiple types of self-governing sensors to co-coordinately monitor a particular going on task. A single sensor has been called 'mote' in this chapter. A mote contains a processor, a sensor and wireless transceiver equipment to do its work. A mote can accumulate the sensor data, perform local processing and transfer the entail information to a base station or cluster head. We are heeding on Crossbow's Mica mote. This mote can support multiple types of sensors and it involves both passive and active sensors. In a typical wireless sensor network, there are hundreds of nodes spread across a particular geographical area for collective monitoring. The rudimentary functionalities of these nodes are detection, categorization and tracking of both static and moving objects. Some of the substantial benefits of these micro sensors are their small-scale size, low power consumption and assist to distributed operation. An explosive is a chemical mixture which can improvise a gigantic explosion usually accompanied by production of light, heat, sound and pressure. As it is a chemical compound, particles of this compound will have their presence in that atmosphere where the explosive occupy. We are also concentrating on the detection of these sorts of particles. Explosives are classified according to the rate at which they stretch. The classifications are high explosives (materials that detonate) and low explosives (materials that deflagrate)

## 3. Architecture

The system consists of a gas sensor, magnetic sensor, a chemical sensor, a TWT Doppler radar unit, a control unit (CU), Wireless Camera, Wireless Antennas and Mica mote. We can also implement this system by embedding it into Traffic Signal, Solar powered road side reflector, Traffic Signals we can embed this in speed breaker to sense the lower part of Vehicle Engine. This proposed system is divided into two parts. First is Control Unit which is used to control and initializes the sensors and do some calculation through the use of cluster head and the other part is software based which is named as Security Alert System that will analyze and monitor the working of proposed system. It also generates different alerts on the basis of calculation received on base station through control units.

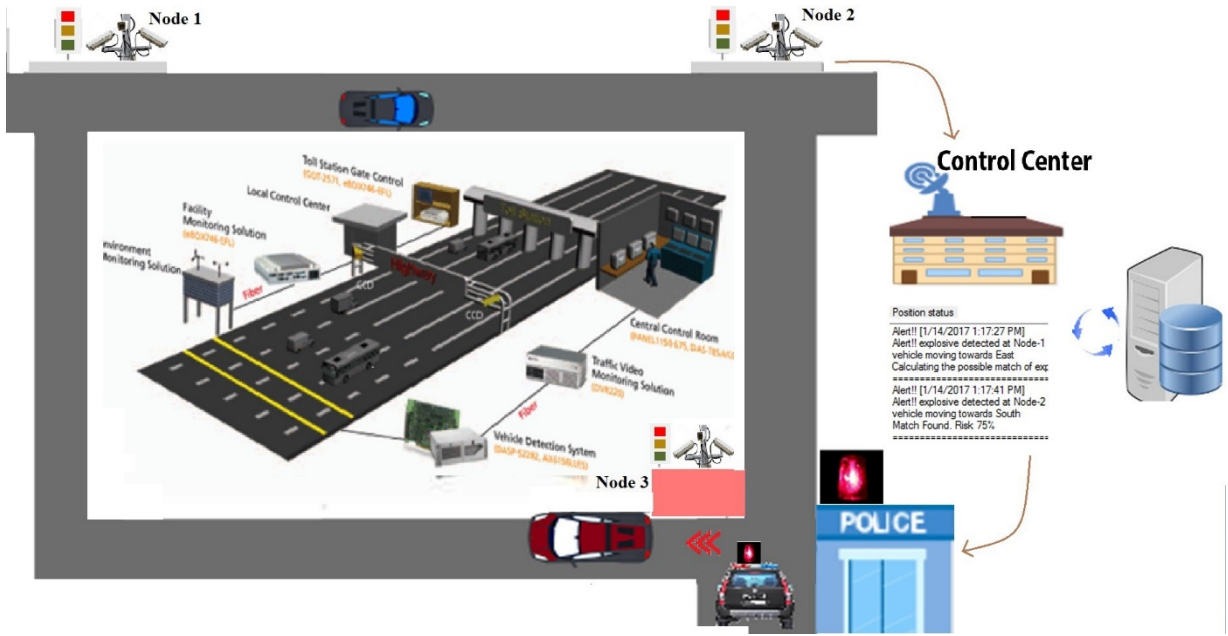

## 3.1 Non-Intrusive Sensor

Non-Intrusive sensors are used to overcome the disadvantages of intrusive sensor. These sensors are placed above at an angle perpendicular to traffic lane. e.g. Video Image Processing (VIP), passive acoustic array, laser and microwave radar.

## 3.2 Smart-Dust Sensor Node

Smart Dust Sensor Nodes are small in size and have low power design. Its hardware contains processor, memory, sensor, radio. There we have three generations of Smart Dust Sensor Node. 1st Generation "Rene Mote", 2nd Generation "Mica Mote", 3rd Generation "Mica Dot Mote". It consists of two major components "Motherboard and Sensor Board". Atmel 90LS8535 processor, 8KB Flash RAM, 512KB SRAM and a RF transceiver for wireless communication is placed on "Motherboard". The Sensor board consists of a 10-bit analog to digital converter, a Magnetometer (Honeywell HMC1002), a temperature sensor, a photo camera and an accelerometer sensor.

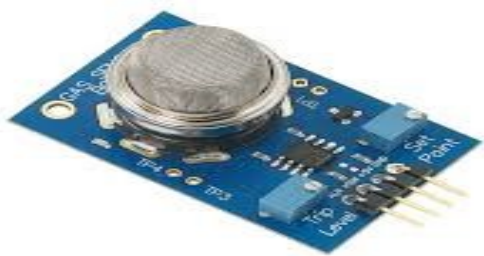 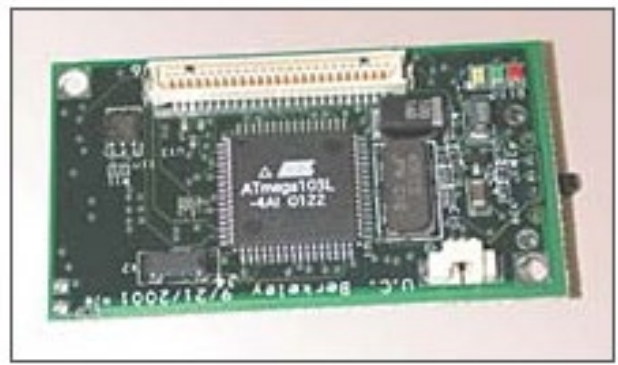

## 3.3 Traffic Camera

These cameras are connected wirelessly via cellular technology. These cameras are used to capture and send High definition pictures or video streams of target object which is considered to be an object containing the explosive material. These Wireless cameras are controlled using IoT (Internet of Things) with WSN through Control Unit to store the images in database or in Base Station.

## 3.4 Spectra Cellular Routers

These are used to increase the speed of data transfer. Like we install these routers above roadways and these are connected directly to traffic cameras to transfer or capture images or video streams of target object at the speed up to 50Mpbs upload and 100Mbps download. These are also connected directly to LED boards over the roadway to automatically and instantly change the information which is displayed to the vehicles moving on the road for controlling the situation.

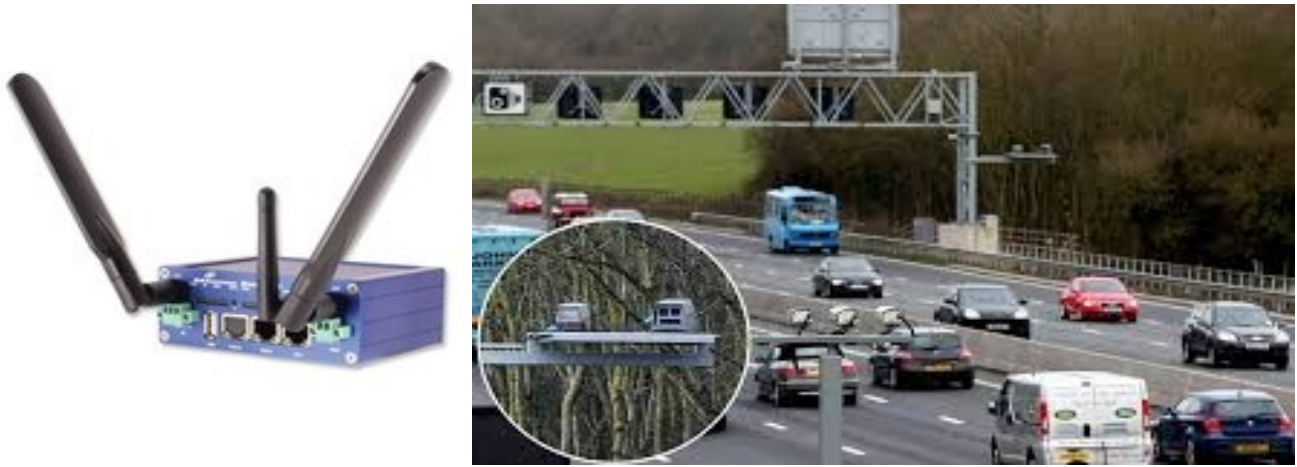

### 3.5 Magnetic Sensors

This sensor measures the strength, direction and magnetic flux of a magnetic field. An input which creates the magnetic field acts to make change in the magnetic field e.g. a ferrous object moves within the earth's magnetic field. For this many sensing devices are used such as Magneto resistive devices (measures electrical resistance as a function of applied magnetics flux) and coil or flux-gate sensors (measures the difference in magnetic field). In this system, we are using Magnetic Sensor for sensing the ferromagnetic material found in explosive while detection like it can be found in any vehicle or in any case or in a container. All information or positive presence of material will be delivered to Control Unit using WSN.

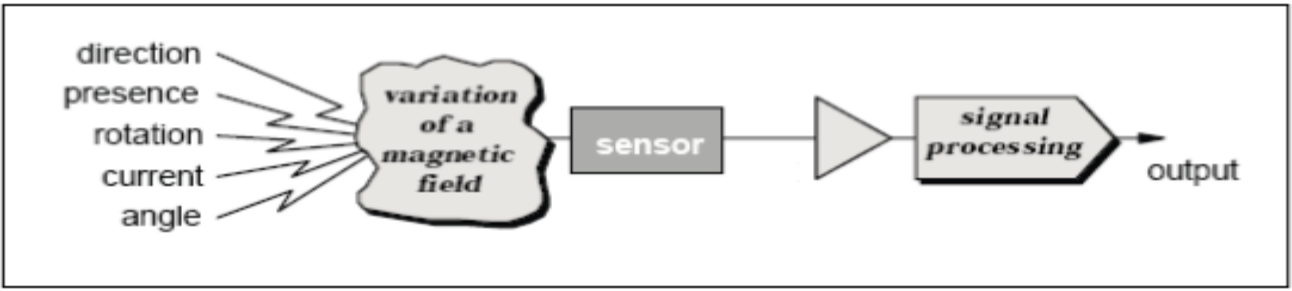

### 3.6 TWT Doppler radar

We are using an onboard radar device TWR-ISM-002 pulse Doppler Sensor. It can sense the mobile vehicle or object up to 30 ft radius. [3] Tracking of the target object is done by this module. It will give periodic updates about the position coordinates of the target object. This will keep track of the target until the target moves out of its range. The function of this Radar is mainly to provide the current position coordinates and forward them to Control Unit and Cluster Head for next estimation and measuring of possible locations where the target can move next using Map to control the traffic and generates alert using IoT phenomenon in Control Unit.

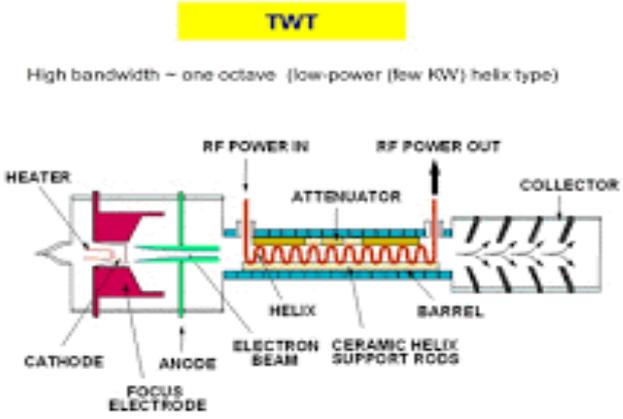

## 3.7 Chemical Sensor (CS)

This sensor is used to determine the chemical signature of the explosive found. As we know different chemical s have different chemical signature. Here, we will be having a database that contains the chemical signatures of almost all explosives. Whenever the control unit found a match it will mark that compound found in the material of target object as an explosive. If a positive presence of explosive found then Gas Sensor Will be invoked by Control Unit to sense the Gas particles.

## 3.8 Gas Sensor (GS)

It will capture the air particles and checks for the presence of explosives in that specific area. If Chemical sensor do not mark for the positive presence of explosive then Gas Sensor has to sense the air particles and material for some time as a particle will remain in the air for 5 minutes after the target object moves. So, after sensing the air if positive presence found then it will mark the explosive and forwards the explosive information to the other nodes in the WSN.

## 3.9 Control Unit (CU)

Control Unit is the central part of the system. CU classifies each explosive found based on their chemical signature, particle concentration, magnetic flux or behavior predicted in ferrous object. Control Unit (CU) receives each and every part of data from other sensor nodes placed in the WSN. Control Unit (CU) contains a memory unit, memory card integrated storage for storing data delivered to it from sensor nodes and a high-speed processor for processing and calculating the results by matching the signatures or other data from database. If it found any match it will mark the target object and send signals to TWT Doppler Radar using IoT phenomenon starts tracking its location and also invokes the traffic cameras which is connected to Routers to generates or captures images of target object. After capturing the images, it will check the identity of target object by searching into database by employing some Face-detection Algorithms. If the target's match found in database then it will mark the object with all its information fields in like retrieve name, retrieve address and decide if the target falls under the category of terrorism. And send all retrieved information to other nodes and forward a copy of information to base station for proper investigation. Control Unit (CU) will perform all necessary computation so it must be deployed with a high processing speed and a reliable hardware unit.

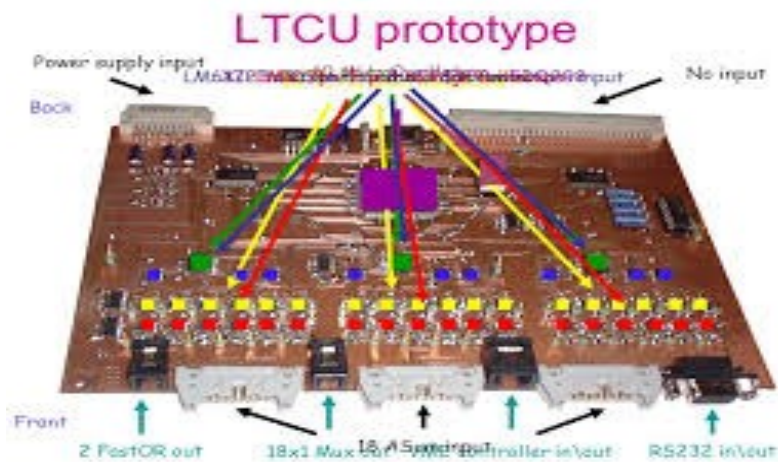

## 4. ALGORITHM

1. Initialize all the nodes in WSN.
2. Synchronize all the nodes and units in the network.
3. To save energy the system will goes to energy saver mode (sleep).
4. Magnetic Sensor sense the presence of any ferromagnetic material like any weapon.
5. If found then it will check for the type of weapon and detect the target object identity and informs the emergency unit to take the target under observation.
6. If Chemical Sensor detects the presence of explosive material.
7. Chemical Sensor will initialize the Gas Sensor to sense the explosive gas if found.
8. Chemical Sensor will also initialize the Control Unit from energy saver mode to active mode.
9. Control Unit stores the reading data of chemical Sensor.
10. Control Unit waits for the Gas Sensor reading.
11. If any ferromagnetic material detected by Magnetic sensor (MS), it will invoke the Control Unit (CU).
12. The control Unit stores the reading of Magnetic Sensor and checks whether the Material is explosive or not.
13. Control Unit initializes the Radar to keep track of the Target Object.
14. If Gas Sensor senses the particle concentration then it will pass the reading data to the Control Unit.
15. Control Unit stores the reading of Gas Sensor.
16. Control Unit uses both readings and checks the signatures of explosive in database.
17. If the Signature of chemical compound matches from database.

18. If unknown chemical compound found then according to characteristics of chemical it will be added to database with a new entry.
19. Control Unit will invoke the Traffic Camera or Video Image Processing (Smart-Dust Sensor for capturing the traffic or object on the lane).
20. Traffic Camera using High Speed Wireless Antenna's will capture the target object.
21. Control Unit takes the picture of the target object and stores it to database.
22. If the target object is a human being then Control Unit (CU) will find a match of object's picture using different face detection algorithms over database.
23. After fetching the target object records/information from country database it will add the target object in brown list in database to keep it under observation.
24. If target object (human) already blacklisted then Control Unit (CU) will alert and notified the nearest cops headquarter.
25. Control Unit will add the target object with the name of the explosive material found in Brown list.
26. Some (or all) of the nodes in a cluster detect the target object and report their data to a cluster head. Note that, "Each cluster has only one cluster head".
27. The control unit will get the current location of the target object.
28. The cluster head collects all target detection information from different nodes to measure the target's actual location.
29. The cluster head uses the computed target object coordinates location from Radar and the previous location coordinates to predict the next location where target object can move to.
30. Sensors that cover the predicted location of target object will be invoked from sleep mode to active mode to form a new cluster to detect target object.
31. Control Unit (CU) in case of emergency will change the traffic signals to prevent vehicles entering in this area.
32. Control Unit (CU) using IoT conveyed a voice broadcast to all the mobile stations (Cellular phone) present around the emergency area using FM-frequency as every mobile station will work on soft. Based frequencies.
33. Control Unit (CU) will generate the emergency alarm and this cluster will be mapped as red zone.
34. After calculating the amount of explosive from the reading of Sensors using Cluster heads, Control Unit will inform the Base Station about the target identity and its location.
35. Base Station will alert all the security Concerned Departments.

In the meanwhile, the complete analysis will be seen and controlled by Base Station Controllers.

## 5. FUTURE PLAN

For future concern, we will look into the practical analysis of our software based implementation by keeping in view the respective security concerns (e.g. data theft, frequency change, traceable network IP, Geo location accuracy) with a detailed instruction about how we implement our system into different areas of a city, some request/response mechanisms for information generation and security checks. Our research will more emphasize and divided into different parts, Traffic signal management, identity detection and Emergency Alerts and Broadcasting Mechanism. Along with that, we are going to adopt the Cloud computing approach for data storage and processing and then we will also design the system on the basis of Fog Computing. After all, we will conclude the difference between both approaches and prove the best fit approach.

## 6. CONCLUSION

*We have proposed an efficient system for standoff explosive detection and security alert system. While comparing with traditional systems our (**e-DASS**) have lot of benefits. The main advantages are due to its miniature size, distributed operation, low power consumption and easy to install implementation. This system is organized in such a way that only security officials know about the presence of the system, control, authentication and security alert base station monitoring and analysis.*